\documentclass{sig-alternate}
\setcopyright{acmcopyright}
\usepackage{url}
\usepackage[utf8]{inputenc}
\usepackage{color}
\usepackage{t1enc}  
\usepackage{graphicx}
\usepackage{color}   
\usepackage{amsmath} 
\usepackage{amsfonts}   
\usepackage{enumerate}  
\usepackage{longtable}
\usepackage{dcolumn}
\usepackage{bm}
\usepackage{epsfig}
\usepackage{subfigure}
\usepackage{wrapfig}
\usepackage{tikz}
\usepackage{pbox}
\usepackage{tabularx}
\usepackage{enumitem}
\usepackage{hyperref}
\usepackage{enumitem}
\setlist{nolistsep}

\usetikzlibrary{arrows,fit}

\newenvironment{myitemize}{\begin{itemize}[leftmargin=*] \setlength{\topsep}{0pt} \setlength{\itemsep}{0pt} \setlength{\parskip}{0pt} \setlength{\parsep}{0pt}}{  \end{itemize} }

\newenvironment{myenumerate}{\begin{enumerate}[leftmargin=*] \setlength{\topsep}{0pt} \setlength{\itemsep}{0pt} \setlength{\parskip}{0pt} \setlength{\parsep}{0pt}}{  \end{enumerate} }

\newcommand\addvmargin[1]{
  \node[fit=(current bounding box),inner ysep=#1,inner xsep=0]{};
}

\CopyrightYear{2017} 
\setcopyright{othergov}
\conferenceinfo{WSDM 2017,}{February 06-10, 2017, Cambridge, United Kingdom}
\isbn{978-1-4503-4675-7/17/02}\acmPrice{\$15.00}
\doi{http://dx.doi.org/10.1145/3018661.3018664}

\begin{document}

\title{Raising Graphs From Randomness to Reveal \\ Information Networks
}

\numberofauthors{1} 
\author{\alignauthor R\'obert P\'alovics$^{1,2}$ \qquad Andr\'{a}s A.\ Bencz\'{u}r$^{1}$ \\
\affaddr{$^1$Institute for Computer Science and Control, Hungarian Academy of Sciences} \\
\affaddr{$^2$Budapest University of Technology and Economics} \\
\email{palovics@sztaki.hu, benczur@sztaki.hu}
}


\maketitle

\begin{abstract}
We analyze the fine-grained connections between the average degree and the power-law degree distribution exponent in growing information networks.
Our starting observation is a power-law degree distribution with a decreasing exponent and increasing average degree as a function of the network size.
Our experiments are based on three Twitter at-mention networks and three more from the Koblenz Network Collection.
We observe that popular network models cannot explain decreasing power-law degree distribution exponent and increasing average degree at the same time.

We propose a  model that is the combination of exponential growth, and a power-law developing network, in which new ``homophily'' edges are continuously added to nodes proportional to their current homophily degree.
Parameters of the average degree growth and the power-law degree distribution exponent functions depend on the ratio of the network growth exponent parameters.
Specifically, we connect the growth of the average degree to the decreasing exponent of the power-law degree distribution.
Prior to our work, only one of the two cases were handled.
Existing models and even their combinations can only reproduce some of our key new observations in growing information networks.
\end{abstract}



\section{Introduction}

While information spread is a main  effect in the worldwide social network, the actual mechanics of this process is especially hard to analyze, as noted among others by Liben-Nowell and Kleinberg~\cite{liben2008tracing}.
Online social networks are widely used for information network analysis~\cite{bakshy2011everyone,cha2010measuring,cha2008characterizing,kwak2010twitter}.
While many online social services exist, it is hard to find data sets that reveal detailed temporal records of the network growth.

The concept of \emph{growth} is a blend of two mechanisms that can be hardly separated.
First, information networks grow  naturally by newly established connections between individuals.
Second, one can view the growth as a network discovery process.
Interactions of users observed via information flow partly reveal the hidden social network of them.

\begin{figure}
\centering
\includegraphics[width=4.1cm]{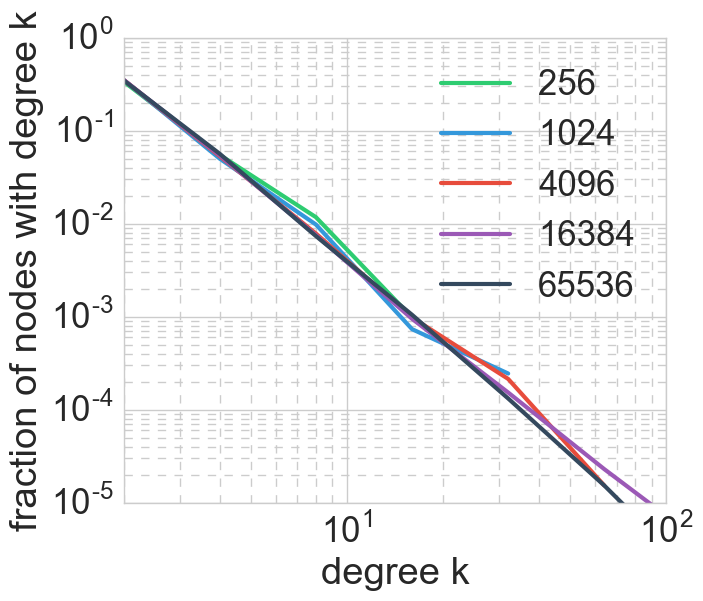}
\includegraphics[width=4.1cm]{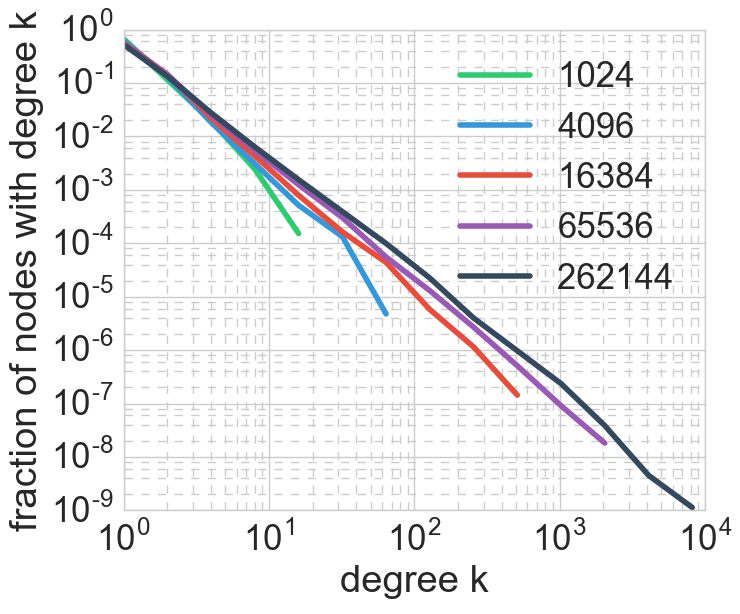}\\
\caption{Degree distribution snapshots of growing networks at different sizes. \textbf{Left:} The Barab\'asi-Albert model yields fixed exponent. \textbf{Right:}  The Occupy data set (see Section~\ref{sec:es-meas}) with flattening slope as the network grows.}
\label{fig:barabasi-vs-real}
\end{figure}

We study the growth of information networks by  considering processes where each node and edge is added to the network only once, and no node or edge is deleted from the network.
Our key finding is related to how the average degree and the power-law degree distribution of the network evolves over time.
More specifically, 
\begin{myitemize}
  \item the exponent of the power-law degree distribution in the network \emph{decreases down to two} over time, and
  \item the average degree grows as  $a + c n ^b$, where $n$ is the number of nodes in the network.
\end{myitemize}
For example, as seen in Fig.~\ref{fig:barabasi-vs-real}, in graphs generated by the Barab\'asi-Albert model \cite{barabasi1999emergence}, the degree distribution exponent stays very close to constant.
The starting point of our observations is that in a real network, the degree distribution log-log plot lines get flattened as the networks grow.

\begin{figure}
\centering
\includegraphics[width=7cm]{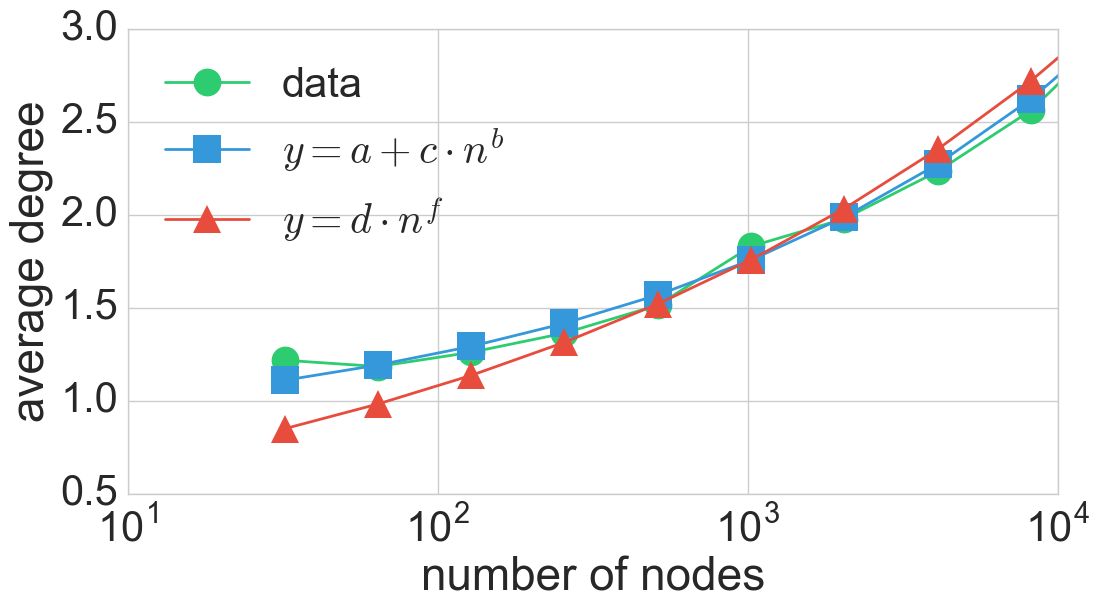}
\caption{Growth of the average degree in the Occupy data set (see Section~\ref{sec:es-meas}).}
\label{fig:leskovec}
\end{figure}

We emphasize the importance of the constant $a$ in the average degree formula. The constant was considered negligible in the experiments of Leskovec et al.\ \cite{leskovec2007graph}. In our results, however, the constant helps capture the mixture of edges that appear at random vs.\ as a result of common interest, and fit to the actual measurements as seen in Fig.~\ref{fig:leskovec}.

To our best knowledge, there is no graph model yet that captures both effects simultaneously.
Leskovec et al.\ \cite{leskovec2007graph} observe densification, a power-law growth for the average degree.
Their models apply to graphs where the exponent of the degree distribution is less than 2 and remains constant over time.
They predict densification for networks with power law exponent larger than 2 that is the case for all of our real networks, however they give no models.

Our exponential growth and preferential attachment based model results in networks with initially slowly growing average degree in addition to decreasing power law exponent.
The main difference of our model compared to earlier models can be summarized in three points.
\begin{myitemize}
  \item The power law exponent, as in all our real networks, is greater than 2 that could not be modeled in \cite{leskovec2007graph}.
  \item Our model explains the initial behavior of the degrees as a natural mixture of influence and preferential attachment edges, also correctly predicting the ratio of these edges.
  \item Our model generates the effects of both increasing average degree and decreasing power law exponent.
\end{myitemize}

\begin{figure}
\centering
\includegraphics[width=5cm]{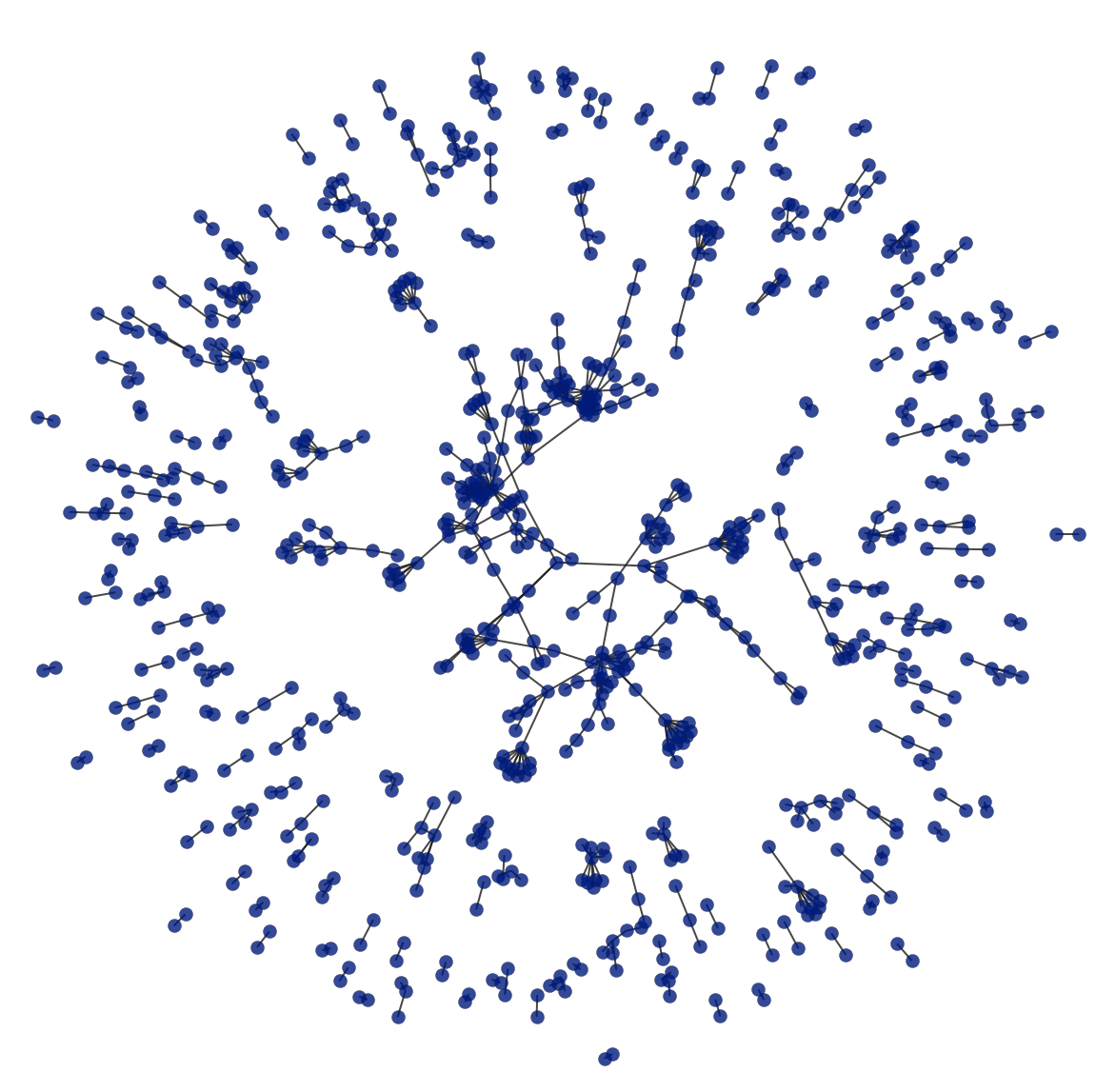}
\caption{The Occupy data set when the number of nodes is 1,000, with several disconnected components, random node pairs, and some high degree nodes. Isolated nodes are not drawn.}
\label{fig:viz_oc}
\end{figure}

As a general overview of the possible models based on our observations, networks start to grow at random, like an Erd\H{o}s-R\'enyi graph.
Then certain rule such as preferential attachment~\cite{barabasi1999emergence} intensifies during the growth process, and causes scale-free degree distribution with a decreasing exponent.
The stronger the rule is, the closer the exponent of the degree distribution gets down to two in a more coherent network.
As the degree distribution log-log plot flattens, the chance for very high degree nodes in a strongly skewed distribution increases that usually act as the main organizer of the network structure.
The intermediate stage in the life of a growing network is illustrated in Figure~\ref{fig:viz_oc}, with the number of nodes around 1,000.
One can observe several disconnected random edges and some larger components with high degree nodes that have already started evolving their internal structure.
As our main result, we aim to model the transition from a random and mostly disconnected graph to a highly organized and very skew degree distribution network.


\subsection{Reproducibility}
\label{sect:reprod}

All data sets used in our measurements are available in an anonymized format at our website\footnote{\small\url{https://dms.sztaki.hu/en/letoltes/networkgrowth}}.
Measurements covered in the next sections are further detailed and illustrated with more figures in Jupyter Notebooks\footnote{\small\url{http://jupyter.org/}} that  are available in our GitHub repository\footnote{\small\url{https://github.com/rpalovics/networkgrowth}}.

\section{Related Results}
\label{sec:related}

Information spread in social networks like Twitter or Facebook have been analyzed in~\cite{bakshy2011everyone,cha2010measuring,cha2008characterizing,kwak2010twitter}.
Several properties of these networks are widely considered to be approximately distributed as power law, with a few exceptions found in \cite{petersen2016power}.
Newman reviews the theoretical background of power law functions and distributions found in empirical data sets in \cite{clauset2009power,newman2005power}.
We rely on these calculations in Section~\ref{sec:es-meas}, where we estimate the exponent of several power-law degree distributions.

A large number of models describing growing networks operate by assuming a constant average degree.
For instance, the model of Barab\'asi et al.~\cite{barabasi1999emergence} is capable of generating networks with power-law distributions, but the average degree of the network remains fixed during the growth of the network.
Another model that assumes constant average degree is the popular copying model~\cite{kumar2000stochastic}.

However, several measurements from the past report that the average degree of a growing network increases over time \cite{barabasi2002evolution,katz2005scale,redner2004citation,vazquez2001statistics,vazquez2002large}.
Some state that the average degree is a \emph{power-law} function of the number of nodes.
This effect has been named \emph{accelerated growth} by Dorogovtsev et al.~\cite{dorogovtsev2002accelerated}, and \emph{densification law} by Leskovec et al.~\cite{leskovec2007graph}.

\textbf{Densification Laws. }
Leskovec et al.\ \cite{leskovec2007graph} mention that the larger the exponent of the densification law, the denser the network over the growing process.
They demonstrate densification law on six different networks, and introduce the community guided attachment model and the forest fire model.
They claim that both can lead to densification law, and give a theoretical explanation for community guided attachment.

Related to our work, they derive a formula connecting the exponent of densification law and the exponent of the power-law degree distribution of the network.
Their model works if both the degree and the densification exponents are constant over time.
They actually predict that if the degree distribution exponent is fixed and larger than 2, then there is no densification and the average degree remains constant.
They observe that for degree distribution exponents larger than 2 and changing over time, densification is possible in the network, however they give no models in this case.

In contrast to \cite{leskovec2007graph}, we find that the increasing average degree is not a simple power-law function of the network size.
We connect this growth to the decreasing power-law degree distribution exponent.  
We propose models that reproduce both increasing average degree and decreasing power law exponent in evolving networks.


\textbf{Work of Dorogovtsev et al.}
In~\cite{dorogovtsev2002accelerated}, the notion of ``accelerated growth'' for power-law degree distribution with increasing average degree is introduced.
Similar to Leskovec et al., their formulas connect the exponent of the degree distribution and the accelerated growth exponent.
The main difference is that they predict densification for networks with fixed degree distribution exponents above 2 as well.
However, in contrast to our key observation, the exponent of the power-law degree distribution remains fixed in all models proposed in~\cite{dorogovtsev2002accelerated}.
Furthermore, we find a different growth for the average degree and connect this to the decreasing power law exponent.

Another work of  Dorogovtsev et al.\ is a growing network model in~\cite{dorogovtsev2000scaling}.
The concept of the model is that while random users continuously connect to the network, edges appear at constant $c$ rate between already existing nodes via preferential attachment.
The model results in a network with power-law degree distribution where the exponent can be derived from the only model parameter $c$ (see ~\cite{dorogovtsev2000scaling}).
A similar model has been analyzed by Barab\'asi et al.\ \cite{barabasi2002evolution}, and by Chung in~\cite{chung2006complex}.
Note that while the model leads to different degree distribution exponents for different values of $c$, for a given growth process $c$ is always fixed in the cited works and the degree distribution remains the same.

\textbf{Models of V\'azquez. }
In contrast to densification law and accelerated growth, in \cite{vazquez2001statistics}  the average degree is measured to increase as a \textit{logarithmic function} in the network size.
V\'azquez investigates network models that generate power-law degree distributions with various exponents and  introduces the triangle closing model.
This dynamic model has only one parameter $u$, and at each time step (i) with probability  $1-u$ we add a new node to the network and connect it to one uniform randomly chosen previous node, and (ii) with probability $u$ we close a possible, previously not existing triangle in the graph.
An in-depth analysis of the model can be found in~\cite{vazquez2003growing}, where the author claims that the process results in a network with power-law degree distribution and the exponent of the distribution depends on the model parameter $u$.
While the model is capable of generating networks with various degree distribution exponents, in a fixed parameter setting, the exponent of the distribution remains constant over time.

\textbf{Random Edge Sampling by Pedarsani et al.}
They question densification law in \cite{pedarsani2008densification} and state that densification can be a result of the fact that measurements on networks are usually carried out on edges sampled from an underlying fixed network.
In other words, they conclude that densification may arise as a feature of the common edge sampling procedure to measure dynamic networks.
They show that network growth can be a direct consequence of the sampling process, therefore the sampling process itself is a plausible explanation of network densification law. In Section~\ref{sec:edge-swap} we introduce a simple experiment indicating that the growth of the average degree can not be explained by random edge sampling.
We simply shuffle our time series that contains timestamped information about the growth of the network. 
We notice that the uniform shuffling procedure yields significantly different average degrees at different sizes of the network.

\textbf{Vertex Copying Models. }
Another popular model that is capable of generating power-law degree distributions is the vertex copying model.
Here at each time step we add a new node that chooses an ambassador node uniform randomly.
Next, we copy each of the connections of the ambassador node to the new node with probability $q$, or attach a new edge to the new node uniform at random with probability $1-q$.
The model leads to a power-law degree distribution with exponent $\alpha = \frac{1-q}{q}$.
Similarly to the models of Dorogovtsev et al.\ and V\'azquez, vertex copying models generate networks with fixed degree distributions.

\textbf{Summary. } 
We can conclude that numerous network models result in power-law degree distributions where the exponent of the distribution is the function of the model parameters.
However, it is common in all of the listed models that for a given parameter setting, the degree distribution remains the same during the growth the network.
Besides the above class of popular models, several papers report power-law growth for the average degree.
The proposed models for this effect miss to capture the fact that while the degree increases, the exponent of the power-law degree distribution decreases in real networks when the exponent is above 2.
Furthermore, in our measurements we report that the average degree grows as $a + c \cdot n ^ b$, and that is different from the densification law.

\section{Power-Law Degree Distributions}
\label{sect:background}

In this section, we summarize the main properties and the measurement of the exponent and the average degree in graphs with power-law degree distributions.
We follow Newman \cite{newman2005power} for properties of power-law distributions and \cite{clauset2009power} for fitting the exponent.
Our main goal is to review and emphasize seemingly simple facts that often become problematic or misleading in experimentation.
We also extend the discussion to prepare the background for our new models.

\subsection{Notations for Degree Distributions}

A graph has power-law degree distribution, if the probability that a given non-zero degree node has degree $k$ in the network, $p(d(i) = k)$ is
$$ p(d(i) = k) = C k ^ {-\alpha}, \hspace{1cm} \alpha>2.$$
If the exponent $\alpha$ is greater than 1, as in all of our example networks, we can compute the value of $C$ by
$$\int_1^\infty p(d(i) = k) dk = 1, \quad C =  1 / \int_1^\infty k ^ {-\alpha} dk = \alpha - 1.$$
The number of degree zero nodes in real networks is less explored in the literature. For simplicity, we introduce the notation NZ for the fraction of nodes with non-zero degree.

\subsection{Relation of Exponent and Average Degree}
\label{sect:avgdeg}


By a simple computation of the average degree in networks with power-law distribution~\cite{clauset2009power},
\begin{equation}
  \resizebox{.9\hsize}{!}{$\overline{d}(n) = \mbox{NZ} \cdot \int_1^\infty (\alpha-1) \cdot k^{-\alpha} \cdot k dk = \mbox{NZ} \frac{\alpha - 1}{\alpha - 2} = \mbox{NZ}  \left [ 1 + \frac{1}{\alpha - 2} \right ].$}
\end{equation}

Let $\Delta:= \alpha - 2$. Note that both NZ and $\Delta$ can depend on the size of the network $n$:
\begin{equation}
  \overline{d}(n) = \mbox{NZ}(n) \left [  1 + 1 / \Delta(n)\right ],
  \label{eq:avgdeg}
\end{equation}
or equivalently
\begin{equation}
  \Delta(n) = 1/ \left [\overline{d}(n)/\mbox{NZ}(n) - 1 \right ].
  \label{eq:delta}
\end{equation}
\eqref{eq:avgdeg} indicates that a size dependent average degree in a network with power-law  degree distribution is caused by
\begin{myitemize}
 \item $\mbox{NZ}(n)$, if the fraction of non-zero degree nodes significantly changes with $n$,
 \item $\Delta(n)$, if the exponent of the degree distribution changes during the growth of the network.
\end{myitemize}

\subsection{Maximum Likelihood Estimate of Exponent and Goodness of Fit}
\label{sect:fit}

Estimating the exponent of a power-law distribution can be problematic in case of shifts in the distribution~\cite{racz2009shift}, or when exponential cutoffs~\cite{newman2005power}  modify the shape of the distribution.  
To overcome these problems, specific estimates have been proposed~\cite{clauset2009power,racz2009shift} and reviewed in \cite{petersen2016power}.
We follow and extend the methodology of~\cite{clauset2009power} by applying the maximum likelihood approximation for the exponent based on the subset of nodes with a degree at least a certain $x$,
\begin{equation}
\hat{\alpha} = 1 + |\{i: x_i \ge x\}| \left [ \sum_{i: x_i \ge x} \ln \frac{x_i}{x} \right ]^{-1}.
\end{equation}
For each $x$, we may define the Kolmogorov-Smirnov statistic KS$(x)$ between the observed distribution $x_i$ and a generated power-law distribution of the corresponding $\hat{\alpha}$.
%

A common method to fit the exponent is to select $x_\text{opt}$ that yields the lowest KS$(x)$.
However, in our experiments in Section~\ref{sec:deg-dist}, the maximum value turns out to be unstable:
for several $x$ values with very different $\hat{\alpha}$, KS is very close to that of $x_\text{opt}$.
For this reason, we will compare three different methods to fit the power law exponent in our experiments.
\begin{myenumerate}
\item We compute the estimate $\hat{\alpha}_\text{all}$ with all $i$, i.e.\ $x = 1$.
\item We select $x_\text{opt}$ that yields minimum KS and use the corresponding $\hat{\alpha}_\text{opt}$.
\item We use the entire set of estimates
\end{myenumerate}
\begin{equation}
  \hat{\alpha}_{0.05} := \lbrace \hat{\alpha} \text{ for $x$ with } |KS(x) - KS(x_\text{opt})| < 0.05)\rbrace.
\label{eq:exponent-confidence}
\end{equation}


\section{Measurements on Real Networks}
\label{sec:es-meas}

In this section, we measure the degree distribution in six real networks.
All these networks are grown in time by processes where each node and edge is added to the network only once, and no node or edge is ever deleted.
Our first task is to measure the connection between the power law exponent, the average degree, and the fraction of nonzero degree nodes as in equations~\eqref{eq:avgdeg} and~\eqref{eq:delta} as the networks grow.
We will also investigate the importance of time in how different types of new nodes and edges may appear in these graphs.


\subsection{Data Sets}

We use three Twitter mention networks and three completely different networks from the KONECT data collection in our experiments.
In Table~\ref{tab:data-edge-sampling} we summarize the sizes of the data sets, by showing for each network the final number of nodes $n$ and the number of edges $e$.

\textbf{Twitter}, a rich information source, lets us define several graphs:
the follower network, the retweet graph formed by retweet cascades, and the mention network.
In our work, we will analyze mention networks defined by authors mentioning other users by their Twitter user names in their tweets.

We collected data sets from Twitter that contain tweets about a certain topic, i.e. contain a given hashtag.
We used the \emph{OccupyWallstreet} and the \emph{15O} tweet collection of~\cite{aragon2013communication}.
Furthermore, we used our own crawl, where we collected tweets about the \emph{Malaysia Airlines Flight 17}.
We started to download tweets immediately after the first news reports.

We defined our graphs by sorting the tweets by time.
If a tweet contained a mention in our data, we added a directed edge to the network between the author and the mentioned user.
Otherwise, we added the author user of the tweet as a new node without any connection to the network.

\textbf{KONECT}
is the Koblenz Network Collection of large network data sets~\cite{kunegis2013konect}.
We selected networks from the collection with timestamped edges evenly distributed in time, and degree distribution close to power law.
For example, we discarded networks that were crawled for several weeks but a significant part of their edges appeared within one day.


\begin{table}
\centering
\resizebox{\columnwidth}{!}{%
\begin{tabular}{|l r |r|r|}
\hline
network                      &  &number of nodes & number of edges  \\ \hline
Twitter Occupy & [\textbf{oc}] & 364,649 & 1,004,961 \\ \hline
Twitter 15O & [\textbf{15}]    & 55,976 & 120,632 \\ \hline
Twitter MH17 & [\textbf{mh}]   & 1,366,691 & 3,489,068 \\ \hline
Facebook wall post & [\textbf{fb}] & 46,947 & 183,335 \\ \hline
DBLP coauthorship & [\textbf{db}] & 1,314,050 & 5,362,414 \\ \hline
Wikipedia & [\textbf{wi}] & 1,870,709 & 36,532,531 \\ \hline
\end{tabular}
}
\caption{Summary of the data sets used.}
\label{tab:data-edge-sampling}
\end{table}

\subsection{Average Degree}
\label{sec:deg}

\begin{figure}
\centering
\includegraphics[width=6cm]{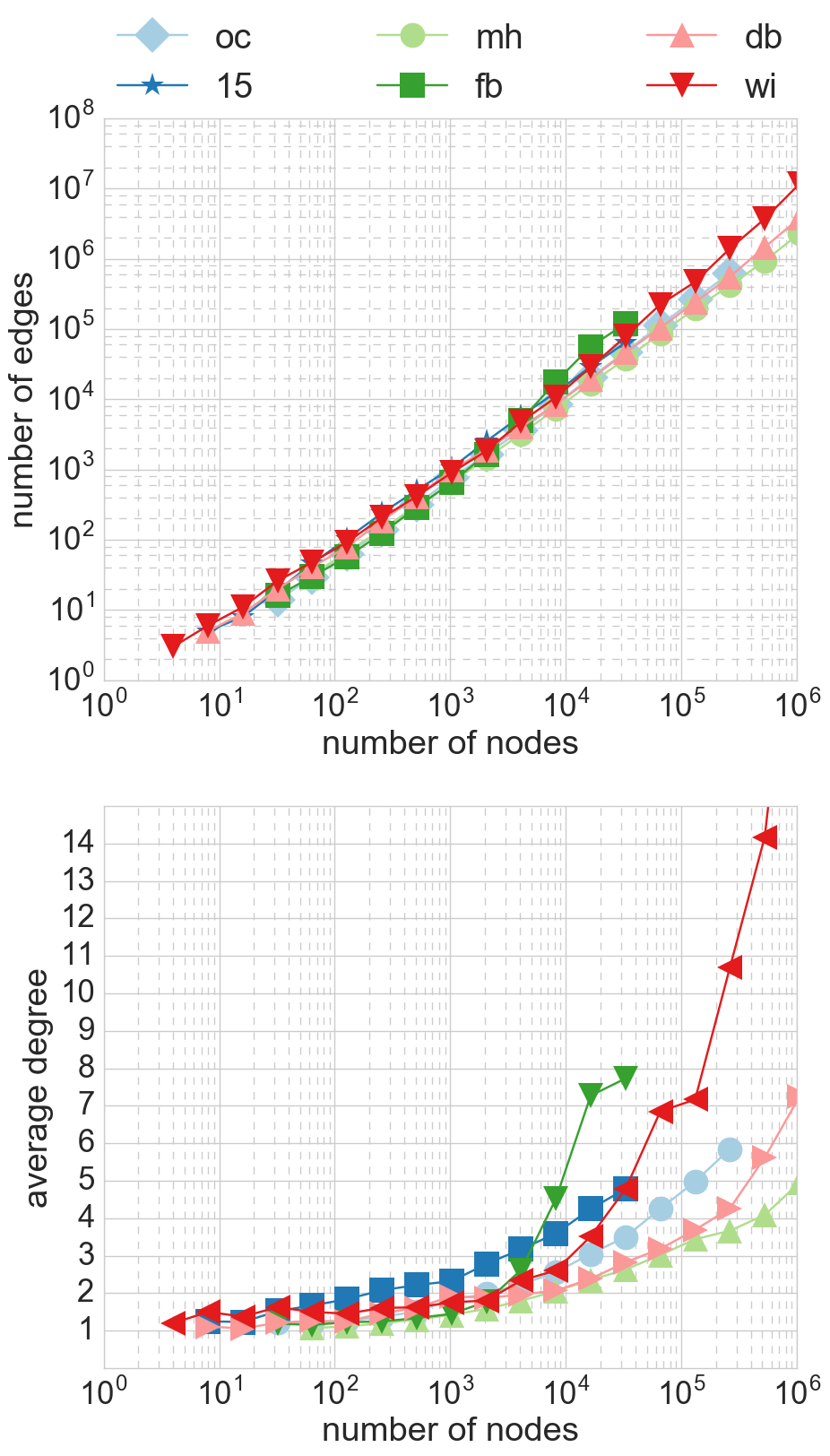}
\caption{Number of edges  (top) and
average degree (bottom) as the function of the number of nodes.}
\label{fig:avgdeg-densification}
\end{figure}

\begin{figure}
\centering
\includegraphics[width=\columnwidth]{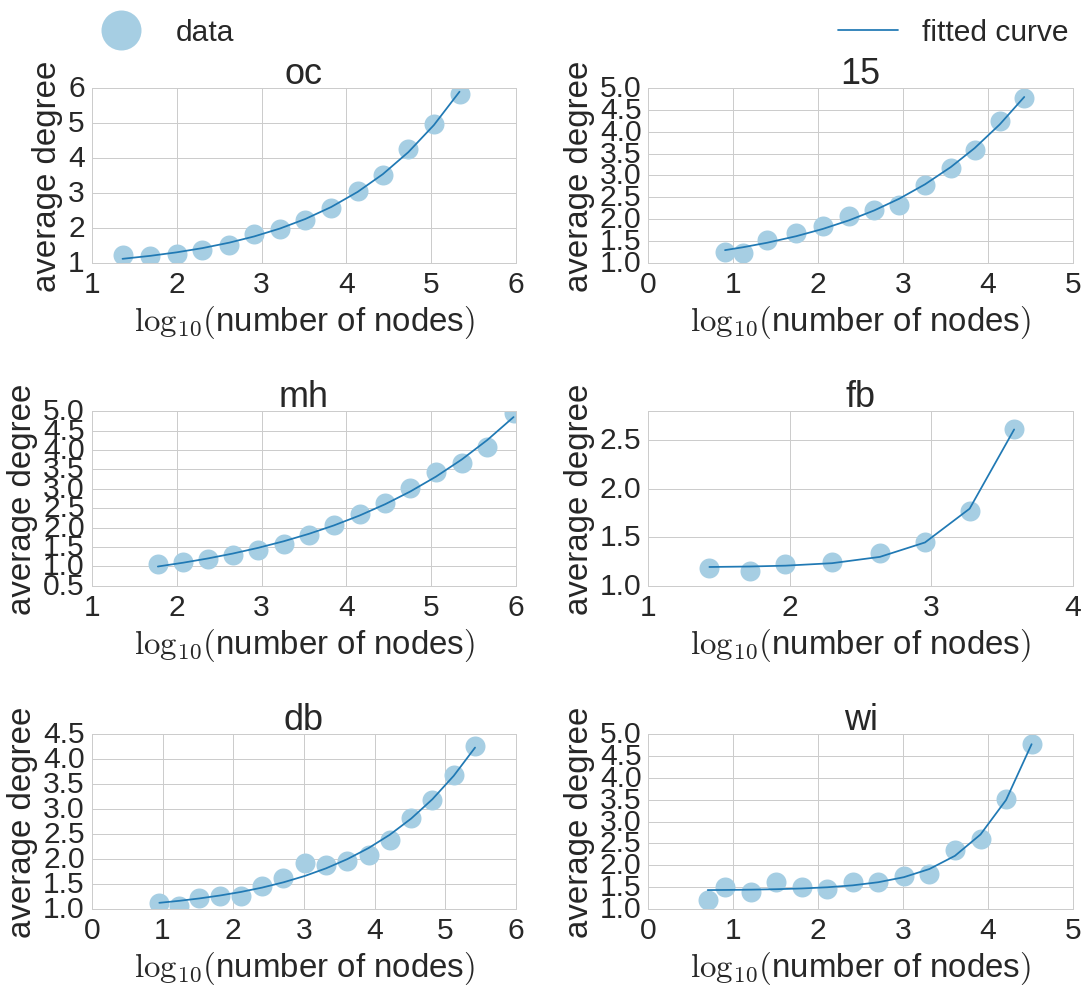}
\caption{Fitted curves for the growth of the average degree for the six different networks.}
\label{fig:fit-densification}
\end{figure}

In our measurements we first measure the number of edges $e$ and the average degree $\overline{d} := 2e/n$ as the network grows, as the function of the number of nodes $n$.
Fig.~\ref{fig:avgdeg-densification} (top) shows the growth of number of edges $e$ for the six different networks in log-log plot.
For all networks, the number of edges increases super-linearly and apparently fit well to a power function, as in~\cite{dorogovtsev2002accelerated,leskovec2007graph}.

Analyzing $e$ as the function of $n$ may however be misleading, since $e$ already grows at least linearly with $n$ and the function will automatically be very close to linear as long as the average degree is small.
Indeed, as seen in Fig.~\ref{fig:avgdeg-densification} (bottom), initially the average degree is around 1, and it slowly starts to increase in all of our networks, which results in a linear growth of $e$ for 3--4 orders of magnitude, without any structural reason.

We find that the increase of the average degree as the function of $n$ can be best modeled by
\begin{equation}
  \label{eq:avgdeg-growth}
  \overline{d}(n) = a + c n^{b} = a + e^{b \cdot \ln(n) + \ln(c)}.
\end{equation}
In Fig.~\ref{fig:fit-densification} we plot the average degrees as the function of $\log_{10} n$.
We show the fit to equation~\eqref{eq:avgdeg-growth} for each network separately.
Our results indicate good fit.
We summarized the parameters of our fitted curves in Table~\ref{tab:fit}, where we indicated the root mean squared error (RMSE) for each fit.

\subsection{Evolving Exponent of Power-law Degree Distributions}
\label{sec:dist}

\begin{figure}
\centering
\includegraphics[width=\columnwidth]{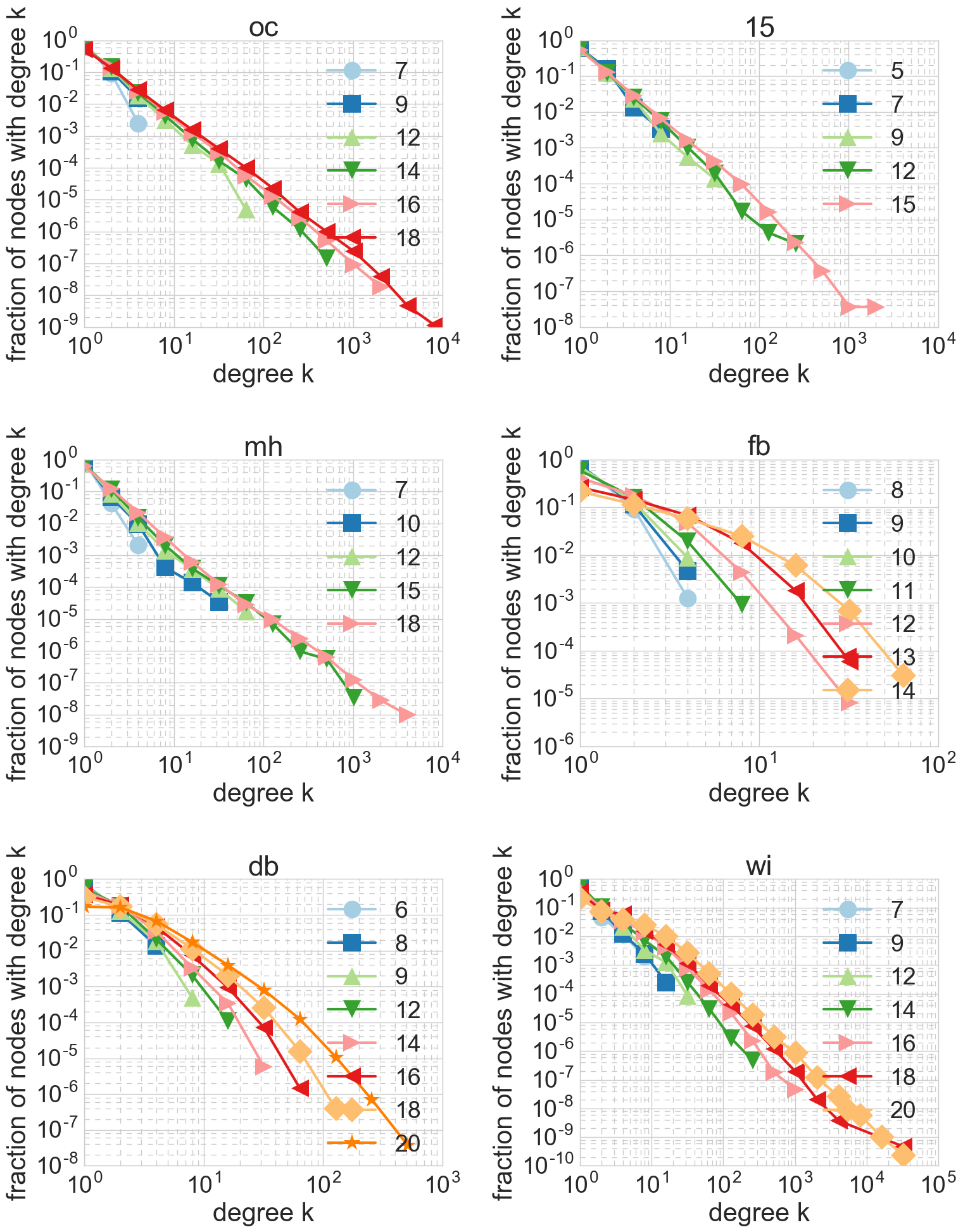}
\caption{Degree distributions for different sizes of the six networks. $\log_2 n$ is indicated in the figures. From left to right and top to bottom: oc, 15, mh, fb, db, wi.}
\label{fig:dists}
\end{figure}

Next we analyze the degree distribution of the observed networks that we compute at exponentially scaled sizes.
In Fig.~\ref{fig:dists}, we plot the degree distribution when $n$ reaches $2^i, i=5,6,7,...$ for the six different networks.
The distributions are exponentially binned and shown on log-log scale.

For all data sets, we may visually observe that \emph{the exponent of the distribution decreases as the network grows} that we will confirm by the methods of Section~\ref{sect:fit} in the next subsection.
We may conclude that the degree distribution changes in time such that the probability of larger degree nodes increases while the power law exponent decreases.

\subsection{Degree Distribution and Average Degree}
\label{sec:deg-dist}

In this section we connect two previous observations, the growth of the average degree and the decrease of the degree distribution exponent.
Recall from Section~\ref{sect:avgdeg} that $\overline{d}(n)$ is the average degree and $\alpha(n)$ is the degree exponent of the network as functions of the number of nodes $n$.
These statistics are connected via the notion $\Delta:= \alpha - 2$ in equations~\eqref{eq:avgdeg} and~\eqref{eq:delta}.
In particular, \eqref{eq:avgdeg}  predicts a connection matching our observation that the average degree grows while the exponent shrinks.
Next we investigate how well the predicted connection between degree and exponent fits to our data.
For our six networks at different sizes, we estimate $\Delta (n) = \alpha (n) - 2$ and $\overline{d}(n)$.

\begin{figure}
\centering
\includegraphics[width=6cm]{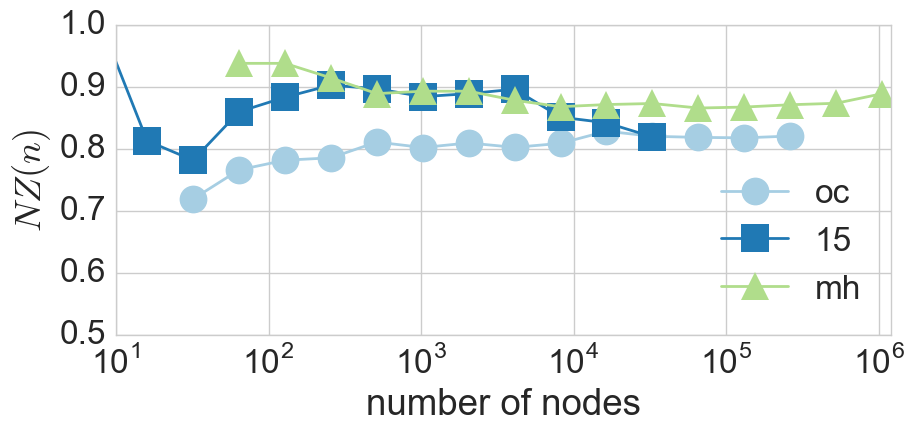}
\caption{Fraction of nonzero nodes NZ$(n)$ for the three Twitter data sets.}
\label{fig:nonzero-twitter}
\end{figure}

We first investigate $\text{NZ} (n)$, the fraction of nodes with non-zero degree in our data, since equation~\eqref{eq:avgdeg} also involves this function.
The networks from the Kob\-lenz collection do not contain zero degree nodes, NZ$(n) = 1$.
The Twitter mention networks include users that have adopted the certain hashtag, but have not been mentioned.
For this reason, NZ$(n)$ is typically less than one for the Twitter data sets, as seen in Figure~\ref{fig:nonzero-twitter}.
Interestingly, NZ$(n)$ is roughly constant in all three cases.

\begin{figure}
\centering
\includegraphics[width=\columnwidth]{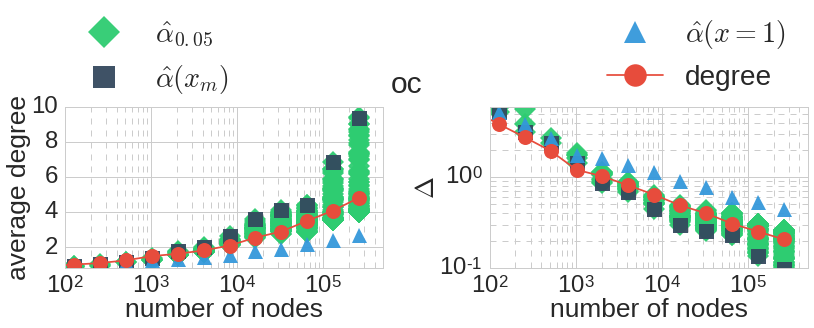}
\includegraphics[width=\columnwidth]{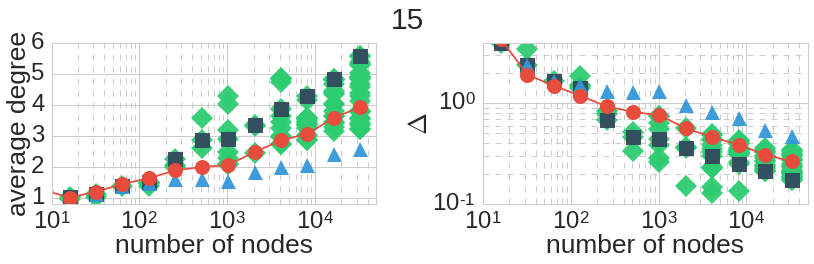}
\includegraphics[width=\columnwidth]{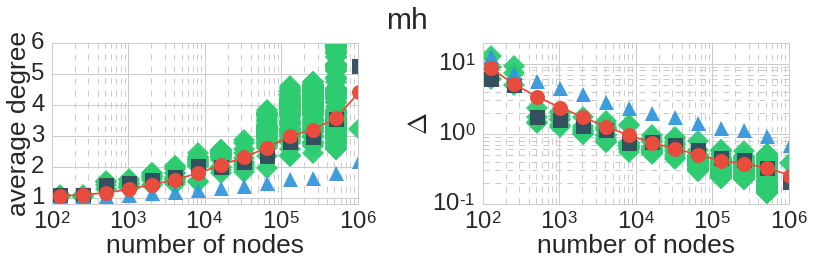}
\includegraphics[width=\columnwidth]{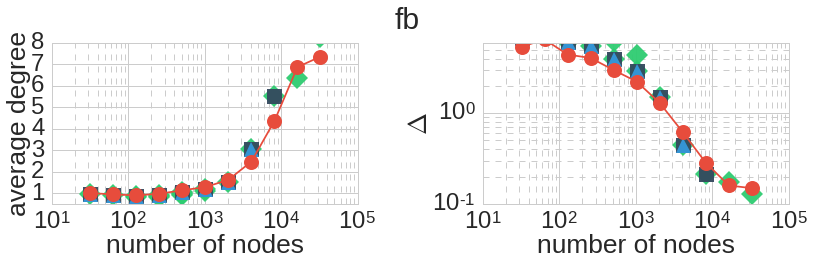}
\includegraphics[width=\columnwidth]{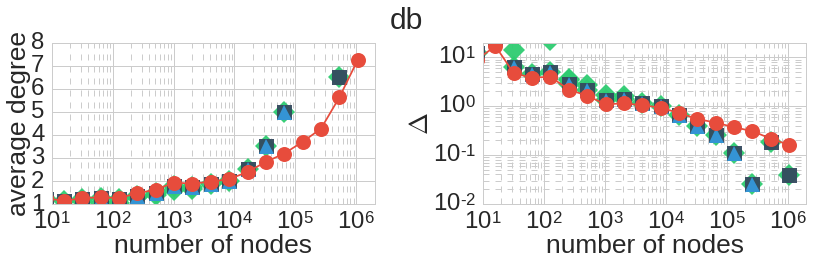}
\includegraphics[width=\columnwidth]{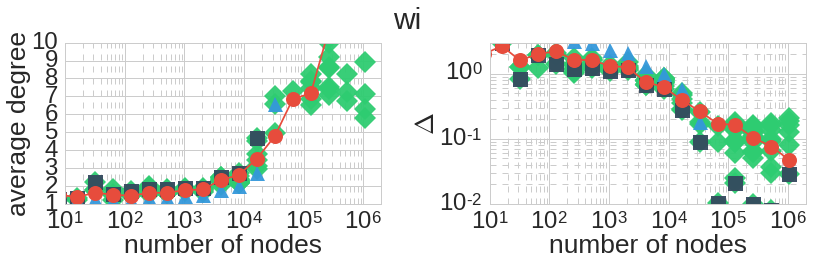}
\caption{\textbf{Left}: Comparison of the average degree (red curve) and its estimates computed from the exponent estimates by using~\eqref{eq:avgdeg}.
  \textbf{Right}: Comparison of the estimates of $\Delta$. The red curve is based on~\eqref{eq:delta}.
}
\label{fig:delta-oc-mh}
\end{figure}

Figure~\ref{fig:delta-oc-mh} shows $\overline{d}(n)$ and $\Delta(n)$ for all the six data sets at exponentially growing size of the networks.
On the left, we show the observed $\overline{d}(n)$, while on the right, the fitted $\Delta(n)$ by the methods of Section~\ref{sect:fit}, i.e.\ the values 
$\hat{\alpha}_\text{all}$, $\hat{\alpha}_\text{opt}$ and the set $\hat{\alpha}_{0.05}$
as in equation~\eqref{eq:exponent-confidence}.
Note that for all networks the exponent is always above 2.

Next, for both sets of plots of in Figure~\ref{fig:delta-oc-mh}, we show the transformation between $\overline{d}(n)$ and $\Delta(n)$.
On the left, we plot the result of equation~\ref{eq:avgdeg} when applied to all the fitted  $\alpha$ values from the corresponding plot on the right. And on the right, we show the result of equation~\eqref{eq:delta} applied to the observed average degree.
The exponent is always above 2, and as it gets closer to 2, $\Delta$ takes very small values shown on logarithmic scale, and $\overline{d}(n)$ increases in the network.
Since the calculation for the average degree tends to infinity as the power law exponent tends to 2 in~\eqref{eq:avgdeg}, the value is highly sensitive to the variation of the fitted exponent when it is close to 2.
Keeping the sensitivity of the estimates near exponent 2 in mind, we observe that the connection between the average degree and the exponent estimates are consistent with the equations~\eqref{eq:avgdeg} and~\eqref{eq:delta}.


\begin{table}
\centering
\begin{tabular}{|r||c|c|c|c|c|}
\hline
~ & $a$ & $b$ & $\ln(c)$ & $c$ & RMSE \\ \hline
\textbf{oc} & 0.8 & 0.29 &  -2.02 & 0.132 & 0.0528 \\ \hline
\textbf{15} & 0.72 & 0.21 & -1.26 & 0.284 &  0.073 \\ \hline
\textbf{mh} & 0.3 & 0.24 & -1.1 & 0.33  & 0.078 \\ \hline
\textbf{fb} & 1.1 & 0.93 & -7.2 & 0.00075 &  0.023 \\ \hline
\textbf{db} & 1.1 & 0.41 & -3.9 & 0.02 &  0.091  \\ \hline
\textbf{wi} & 1.4 & 0.6 & -5.4 & 0.00452 &  0.099 \\ \hline
\end{tabular}
\caption{Parameters of the fitted $a + c \cdot n^b$ curves for the average degree in case of the six data sets.}
\label{tab:fit}
\end{table}

\begin{table}
\centering
\begin{tabular}{ r r c c  }
~ & ~ & $\Delta n $ & $\Delta e$ \\ \hline
Z &\begin{tikzpicture}[-,>=stealth',shorten >=1pt,auto,node distance=0cm,main node/.style={circle,fill=orange,fill}]
  \node[main node] (1) {};
  \addvmargin{0.2mm}
\end{tikzpicture} &  1 & 0 \\  
R & \begin{tikzpicture}[-,>=stealth',shorten >=1pt,auto,node distance=0.1\columnwidth,main node/.style={circle,fill=orange,fill}]
  \node[main node] (1) {};
  \node[main node] (2) [right of=1] {};
  \path[-,very thick,orange] (1) edge node [left] {} (2);
  \addvmargin{0.2mm}
\end{tikzpicture}  & 2 & 1\\  
I & \begin{tikzpicture}[->,>=stealth',shorten >=1pt,auto,node distance=0.1\columnwidth,old node/.style={circle,fill=black,fill},new node/.style={circle,fill=orange,fill}]
  \node[old node] (1) {};
  \node[new node] (2) [right of=1] {};
  \path[orange,very thick,-] (1) edge node [left] {} (2);
   \addvmargin{0.2mm}
\end{tikzpicture} &  1 & 1 \\  
H & \begin{tikzpicture}[-,>=stealth',shorten >=1pt,auto,node distance=0.1\columnwidth,main node/.style={circle,fill=black,fill}]
  \node[main node] (1) {};
  \node[main node] (2) [right of=1] {};
  \path[-,very thick,orange] (1) edge node [left] {} (2);
  \addvmargin{0.2mm}
\end{tikzpicture}  & 0 & 1\\  
\end{tabular}
\caption{Growth events. For each event, the orange part is added to the network.}
\label{tab:growth-events}
\end{table}

As the main conclusion of the measurements in Sections~\ref{sec:deg}--\ref{sec:deg-dist},  in all of the six data sets we observe that
\begin{myitemize}
  \item the exponent of the power-law degree distribution decreases -- i.e.\ nodes with higher degree appear more likely -- but always stay above 2;
  \item the average degree grows as $a + c \cdot n ^ b$;
  \item the connection between the growing average degree and the shrinking exponent, up to the uncertainty in the estimation of the exponent, is as predicted by the power-law distribution.
\end{myitemize}

\subsection{Edge Formation: From Random to Joint Interest}
\label{sec:micro}

The growth of the average degree could be a natural result of new events that generate edges already having one or both endpoints in the network.  Initially, while the network is very small, edges that connect two nodes will likely appear between nodes that have not yet been part of the network, hence the number of edges will increase by one but at the same time, the number of nodes may increase by one or even by two.
Later however, when many of the eventual nodes have already joined the network, the events will bring an extra edge but no new nodes, thus increasing the average degree. 

In order to characterize events with one or two endpoints already in the network, we distinguish four different events in the process as detailed in Table~\ref{tab:growth-events}.
First, a new root node can appear without any connection to the already existing part to the network ($Z$).
Second, a new \emph{random edge} may appear with nodes that are both new in the network ($R$).
Third, one node can appear and immediately join with a new edge to the network ($I$).
This event may be a result of influence between the nodes.
Finally, two existing nodes can connect to each other as a result of common interest, possibly also involving preferential attachment that we call \emph{homophily edges} ($H$).

\begin{figure}
\centering
\includegraphics[width=\columnwidth]{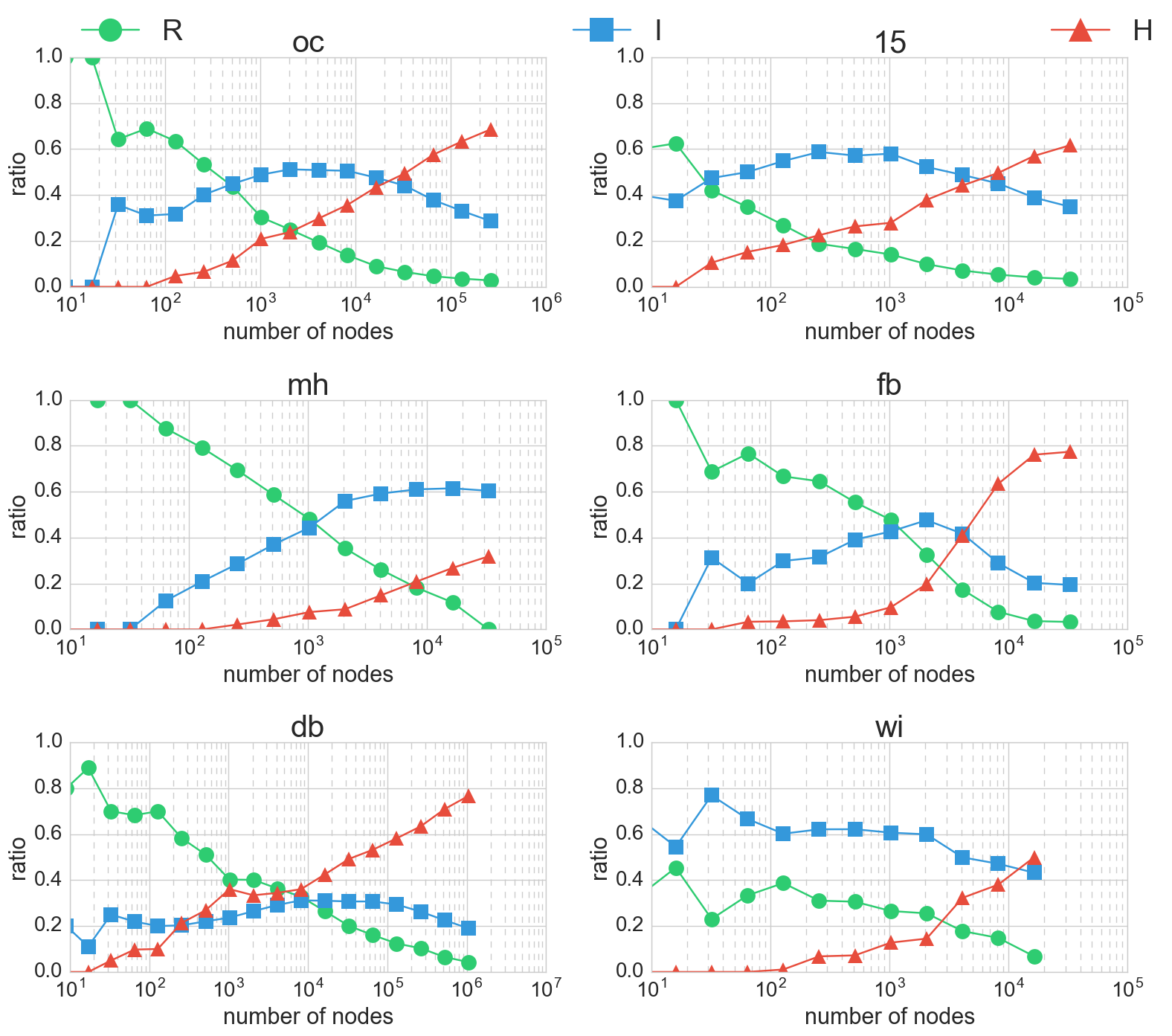}
\caption{Ratio of different types of edges as the function of $n$. $R$, $I$, and $H$ are defined in Table~\ref{tab:growth-events}.}
\label{fig:edge-ratio}
\end{figure}

To measure the increase in average degree as the possible result of increased homophily events \textit{H}, we measure the ratio of different edge types ($R,I,H$) as the function of the network size.
The plots for the six data sets are in Figure~\ref{fig:edge-ratio}.
The most important observation is that the ratio of random edges $R$ declines, and influence events $I$ are roughly constant, and the ratio of $H$ increases as the network grows.

\subsection{No Fit to Uniform Edge Sampling}
\label{sec:edge-swap}

\begin{figure}
\centering
\includegraphics[width=4cm]{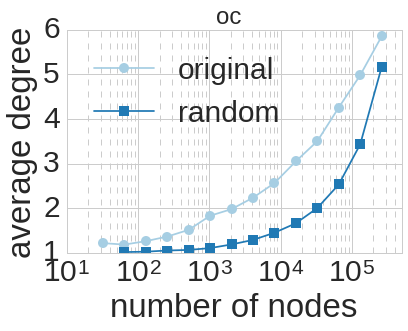}
\includegraphics[width=4.3cm]{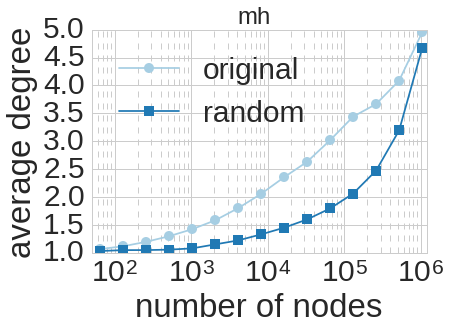}
\caption{Result of shuffling the events in the time series of the Occupy (left) and MH17 (right) data sets. The original graph is always denser than the randomized one.}
\label{fig:edge-swap}
\end{figure}

Pedarsani et al.~\cite{pedarsani2008densification} state that the growth of the average degree can be a result of uniform edge sampling from a hidden network.
We made a simple experiment to investigate their assumptions.
We randomized the edge sequence of the network, and measured the growth of the average degree in the randomized time series.
In other words, instead of observing the edges in a temporal order according to their creation timestamps, we processed them in a random order.
In Figure~\ref{fig:edge-swap} we compare the growth of $\overline{d}(n)$ for the randomized and the original time series in the Occupy (left) and the MH17 (right) data sets.
The results, accessible as described in Section~\ref{sect:reprod}, are similar for the remaining data sets.

According to our measurements, the growth of the average degree cannot be completely explained by random sampling from a hidden social web.
While the average degree generally grows in both cases, it is always larger in the original time series than in the randomized one.
In other words, the actual growth of the network always yields denser graphs than random sampling from the final snapshot of the network.

\section{Modeling}
\label{sec:edge-models}

We showed in Section~\ref{sec:deg} that the average degree increases in evolving networks.
Next, in Section~\ref{sec:dist} we saw that the exponent of the degree distribution of a growing network decreases.
In Section~\ref{sec:deg-dist} we connected the two observations, and concluded that while the evolving exponent of the power-law degree distribution decreases, the average degree grows.
Finally we demonstrated in Section~\ref{sec:micro} that as more and more edges appear between existing nodes, naturally the average degree increases.
We intend to give network models that capture these effects simultaneously,
i.e.\ give a dynamic network model with the following requirements:
\begin{myitemize}
 \item increasing average degree according to $a + c n ^b$,
 \item decreasing power law exponent that tends to 2,
 \item vanishing fraction of random edges,
 \item increasing number of homophily edges.
\end{myitemize}

As far as we know, there exists no network model yet that is able to describe all our observations.
Models that explain increasing average degree are often called densification laws \cite{leskovec2007graph}.
In Section~\ref{sec:model-densification} we will review how existing densification law models predict our other three observations. 
Then in Section~\ref{sec:model-pref} we detail the model of Dorogovtsev et al.\ \cite{dorogovtsev2000scaling} that is closest to our new set of models.
In Section~\ref{sec:model-new}, we introduce our basic model of growing networks with decreasing power-law degree distribution.
After extending the model in~\ref{sec:model-new-extend}, we verify our models with simulations in Section~\ref{sec:model-new-simulation}.

\subsection{Relation to Densification Laws}
\label{sec:model-densification}

Models given by~\cite{dorogovtsev2002accelerated,leskovec2007graph} for accelerated growth and densification law are for graphs with fixed degree distribution exponents.
Moreover, both of them predict for the average degree power-law growth, $\overline{d}(n) =  c n ^b$.
As stated above, in our experiments we measured that the average degree first slowly grows from a constant value that is around 1.
Furthermore, none of the proposed models for accelerated growth and for densification are capable of generating power-law degree distribution with a decreasing exponent.
Finally, in Section~\ref{sec:edge-swap} we showed that the simple uniform edge sampling proposed by Pedarsani et al.\ \cite{pedarsani2008densification} leads to a different network growth process.

\subsection{The Concept of Preferential Edges}
\label{sec:model-pref}

Closest to our result is a growing network model of Dorogovtsev et al.~\cite{dorogovtsev2000scaling}.
The model is capable of generating networks with power-law degree distribution.
In this generative model one part of the edges are added between already existing preferentially chosen nodes in the network.
More specifically, at each time we perform two steps.
\begin{myitemize}
\item A new node is added to the network with $c$ new edges.
The endpoints of the edges are selected at random with preferential attachment.
If the number of edges is $e$ in the network, then the probability that node $i$ with degree $d_i$ gets connected to the new node is $d_i / 2e$.
\item $c$ new edges are added between already existing nodes. If nodes $i$ and $j$ have degrees $d_i$ and $d_j$, the probability of an edge between them is proportional to $d_i d_j$.
\end{myitemize}

The model results in a network with power law exponent, whose value can be derived from a single model parameter $c$, see~\cite{dorogovtsev2000scaling}. 
Since for a fixed $c$, we obtain a fixed exponent, the model cannot explain the shrinking phenomenon.
Similar models have been analyzed by Barab\'asi et al.\ \cite{barabasi2002evolution}, and by Chung in~\cite{chung2006complex}, both of which yield a constant exponent.

\subsection{Model I: Random and Homophily Edges}
\label{sec:model-new}

\begin{figure}
\centering
\includegraphics[width=5cm]{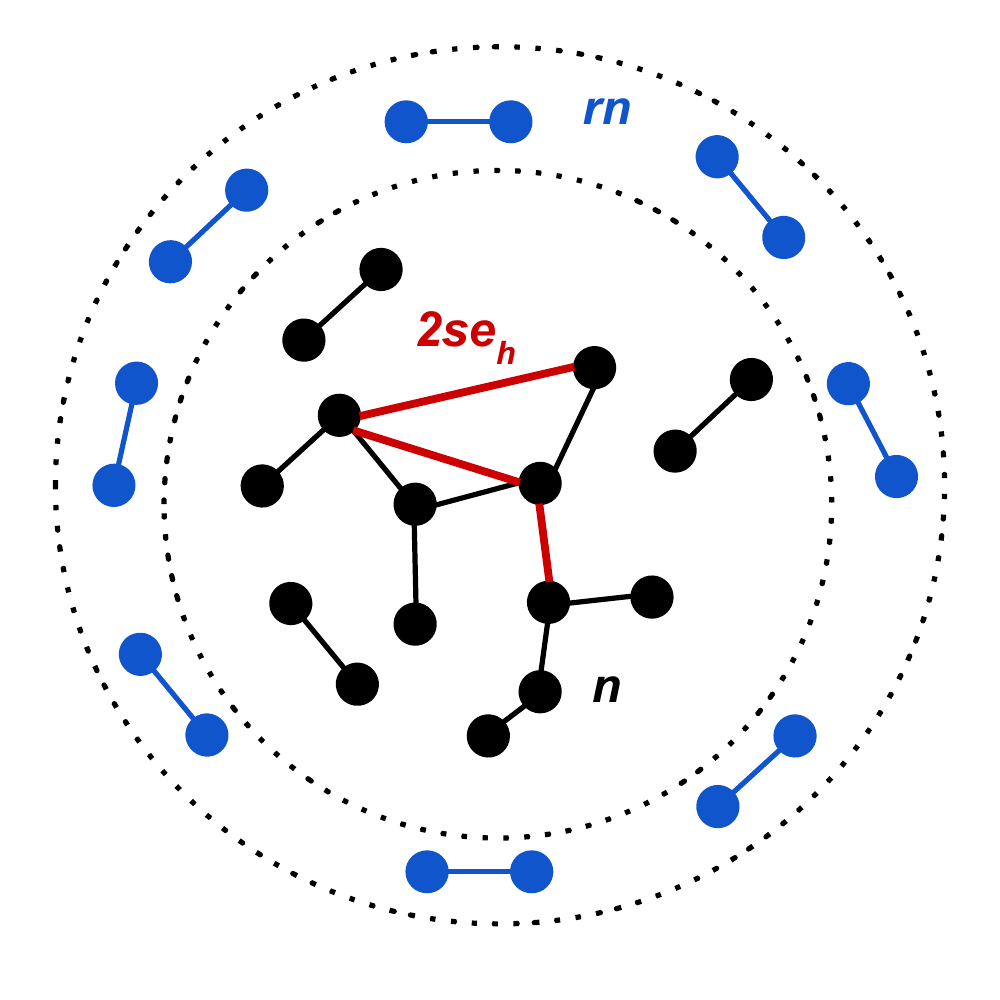}
\caption{Visualization of Model I. At each time $t$, we add $rn(t)$ random, and $2se_h(t)$ preferential edges.}
\label{fig:model1-viz}
\end{figure}

Next we introduce our new model that relies on preferential attachment, and can generate growing networks with decreasing power law exponents.
In our first growing network model we add at each time step (i) random new edges that connect two new nodes in the network, (ii) and edges between already existing nodes in the network. 
More specifically, as indicated in Figure~\ref{fig:model1-viz}, at time $t$: 
\begin{myitemize}
\item For some constant $r$, $r \cdot  n(t)$ new \emph{random edges} appear that indicate the random growth of the network.
\item Each node $i$ selects other nodes to connect with \emph{homophily edges} randomly.
      The expected number of new homophily connections created by node $i$ is $s \cdot d_h(i)$, where
      $d_h(i)$ is the homophily degree of node $i$, i.e. the number of homophily edges connected to node $i$.
      For a given new connection of node $i$, the target node is selected by preferential attachment.
      In other words, the probability of selection for node $j$ as a new neighbor of $i$ is $d(j)$. 
\end{myitemize}
First we analyze the growth of the average degree in the model.
The number of nodes $n(t)$ in the network at time $t$ can be derived from the number of created random new edges by time $t$.
Each new random edge brings in two new nodes, therefore
\begin{equation}
  \frac{dn}{dt} = 2 r n, \hspace{1cm} n(0) = N_0, \hspace{1cm} n(t) = N_0 e^{2rt},
\end{equation}
where $N_0$ is the initial number of nodes in the network.
The number of generated random edges $e_r(t)$ is half of the number of nodes,
\begin{equation}
  e_r(t) = 1/2 \cdot n(t) = N_0 / 2 \cdot e^{2rt}.
\end{equation}

The number of preferentially added edges $e_h(t)$ can be derived from a separated equation.
At each time step the expected number of preferentially crated edges by node $i$ is $ s \cdot d_h(i)$, therefore
\begin{equation}
  \frac{de_h}{dt} = \displaystyle\sum_i s d_h(i) = 2 s e_h,   e_h(0) = H_0,   e_h(t) = H_0 e^{2st},
\end{equation}
where $H_0$ is the initial number of the homophily edges.

As the number of edges $e(t) = e_h(t) + e_r(t)$, the average $\overline{d}(t)$ is then 
\begin{gather}
  \overline{d}(t) = \frac{2 \left (e_r(t) + e_h(t) \right )}{n(t)} = \frac{ N_0 e^{2rt} + 2 H_0 e^{2st} } {  N_0 e^{2rt} } \\
   = 1 + \frac{2 H_0}{N_0} e^{2(s-r)t}  = 1 + \frac{2 H_0}{N_0^{s/r}} n(t) ^ {\frac{s}{r} - 1}
   \label{eq:model1-avgdeg}
\end{gather}

Note that the model generates a growing average degree that increases from constant 1. The exponent of the power-law second term is $s/r - 1$ that is larger than 0 if $s>r$.

The fraction of added random edges is
\begin{equation}
  \frac{e_r(t)}{e_r(t) + e_h(t)} = \frac{1}{1 +  e_h(t)/ e_r(t)} = \frac{1}{1 + 2H_0/N_0 e^{\frac{s}{r} t}}.
  \label{eq:random}
\end{equation}
If $s>r$, that tends to 0 as $t \rightarrow \infty$. Similarly, the fraction of homophily edges is specified by the sigmoid function,
\begin{equation}
  \frac{e_h(t)}{e_r(t) + e_h(t)} = \frac{1}{1 + N_0/(2H_0) e^{\frac{r}{s} t}}
   \label{eq:homo}
\end{equation}
that tends to 1 as $t \rightarrow \infty$. \eqref{eq:random} and \eqref{eq:homo} are consistent with our measurements in Section~\ref{sec:micro} where we observed that the number of randomly appearing $R$ edges decrease as the network grows while the number of homophily $H$ edges increases over time. 

In our model, at the beginning of the growth random edges dominate and the network is similar to an Erd\H{o}s-R\'enyi graph.
Then, the added preferential edges slowly start to dominate the process as computed in~\eqref{eq:homo}.
This effect pushes the network towards a scale-free graph.
The model of Dorogovtsev et al.~\cite{dorogovtsev2000scaling} adds preferential edges between already existing nodes at constant rate $c$ at each time step.
Their model results in power-law degree distribution with a fixed exponent.
In our model as the number of preferentially added  edges increase during the growth, the exponent will start to decrease.
We verify this assumption that is based on the derivations of~\cite{dorogovtsev2000scaling} by simulations in Section~\ref{sec:model-new-simulation}.

\subsection{Model II: Extension with Influenced and Root Nodes}
\label{sec:model-new-extend}

In this section we extend our model of random and homophily edges by introducing two additional events.
\begin{myitemize}
\item We add $p \cdot n(t)$ random influenced nodes that newly connect to the network with one edge. For a new influenced node, the influencer is selected uniform random from the already existing nodes in the network.
\item We add $q \cdot n(t)$ random root nodes that appear in the network with 0 degree.
\end{myitemize}
With similar notations, the number of nodes is
\begin{equation}
  \frac{dn}{dt} = (p + q + 2 r ) n, \hspace{0.2cm} n(0) = N_0, \hspace{0.2cm} n(t) = N_0 e^{(p + q + 2 r)t}.
\end{equation}
The number of influenced nodes $n_i(t)$, the number of zero degree root nodes $n_z(t)$, and the number of nodes added by random edges $n_r(t)$ are then
\begin{gather}
n_i(t) = \frac{p}{p + q + 2 r} N_0 e^{(p + q + 2r)t}, \\
n_z(t) = \frac{q}{p + q + 2 r} N_0 e^{(p + q + 2r)t}, \\
n_r(t) = \frac{2 r}{p + q + 2 r} N_0 e^{(p + q + 2r)t}.
\end{gather}
Hence the number of influence edges, root edges and random edges are
\begin{gather}
e_i(t) = n_i(t), \hspace{0.3cm} e_z(t) = 0, \hspace{0.3cm} e_r(t) = 1/2 n_r(t). 
\end{gather}
The differential equation for the number of homophily edges remains the same,
\begin{equation}
  \frac{de_h}{dt} = \displaystyle\sum_i s d_h(i) = 2 s e_h, e_h(0) = H_0, e_h(t) = H_0 e^{2st}
\end{equation}
The average degree based on the previous formulas is
\begin{gather}
  \overline{d}(t) = \frac{2 \left (e_r(t) + e_i(t) + e_h(t) \right )}{n(t)} \\
   = \frac{ 2 \frac{r + p}{p + q + 2 r} N_0 e^{(p + q + 2 r)t} + 2 H_0 e^{2st} } {  N_0 e^{(p + q + 2 r)t} } \\
   = \frac{2(r + p)}{p + q + 2 r} + \frac{2 H_0}{N_0} e^{(2s-p-q-2r)t} \\ 
   = \frac{2(r + p)}{p + q + 2 r} + \frac{2 H_0}{N_0^{2s/(p+q+2r)}} n(t) ^ {\frac{2s}{p+q+2r} - 1}.
\end{gather}

Note that the process generates a growing network where the average degree grows by $\overline{d}(n) = a + c \cdot n ^ b$, where
\begin{myitemize}
  \item $a = \frac{2r + 2p}{p + q + 2 r}$. In other words, the fraction of nodes added by influence $p$, the fraction of nodes added as root $q$, and the fraction of nodes added by random edges $r$ sets together the constant term.
  \item $b = \frac{2s}{p+q+2r} - 1$ is the exponent of the power-law growth.
  \item $c = \frac{2 H_0}{N_0^{2s/(p+q+2r)}}$ is the multiplier of the power-law term.
\end{myitemize}

The model is capable of generating networks with isolated nodes as in our Twitter data sets.
In case of networks from the Koblenz Network Collection, the average degree is always greater than 1, and there are no root nodes in the growing network.
By setting the proper fraction of random edges $r$ and influenced nodes $p$, and setting $q=0$, the model is also capable of modeling these networks.

\subsection{Verifying the Models by Simulations}
\label{sec:model-new-simulation}

\begin{figure}
\centering
\includegraphics[width=4.55cm]{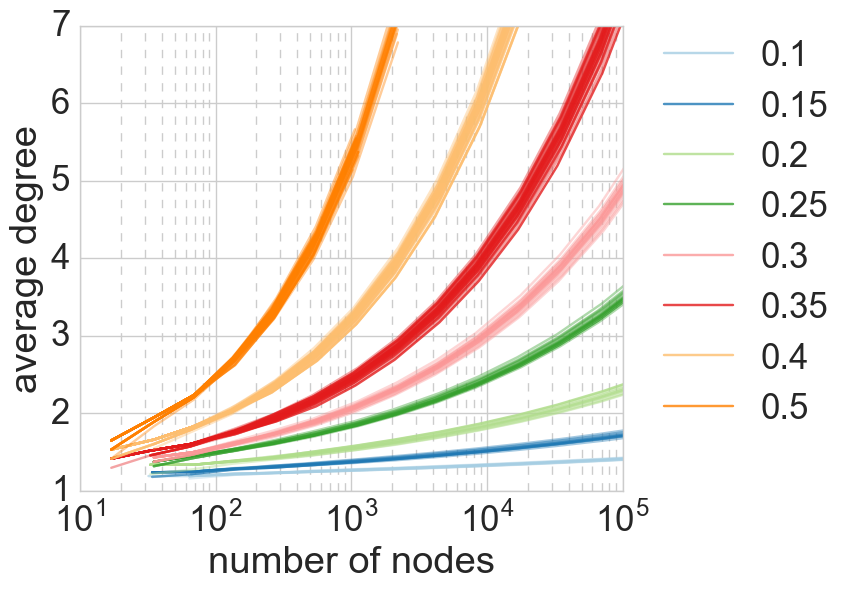}
\includegraphics[width=3.65cm]{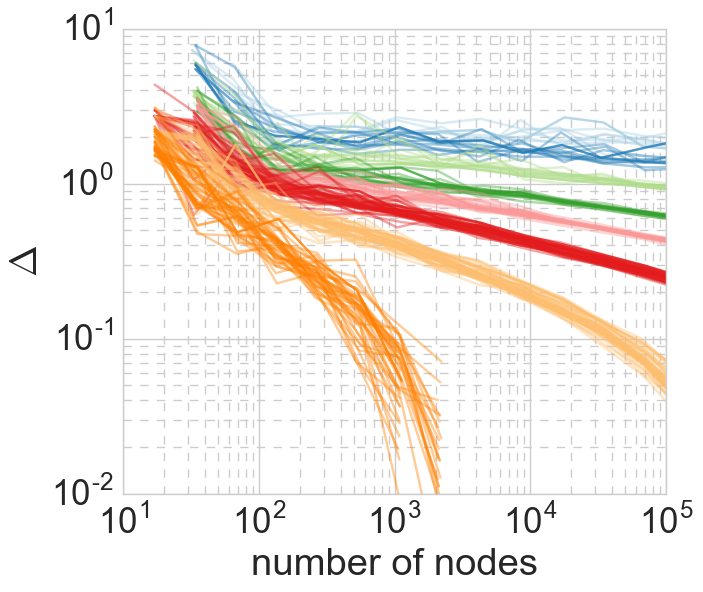}\\
\caption{Results of simulations for Model I. Different values of $s$ correspond to different colors as noted above. \textbf{Left:} $\overline{d}(n)$ \textbf{Right:} $\Delta(n)$.
\label{fig:model1-curves}
}
\end{figure}

First we investigate the properties of networks generated by  Model~I of Section~\ref{sec:model-new}.
In Fig.~\ref{fig:model1-curves} we fixed $H_0 = 2, r = 0.05$, and ran simulations at different homophily rates $s$.
On the left, we show the increase of the average degree, while on the right, the estimate of $\Delta$ while growing the simulated network.
Note that different colors correspond to different values of $s$, and we ran 40 different simulations in case of each parameter setting.
The results indicate that $s$ sets the velocity of the growth of $\overline{d}(n)$.
Furthermore, $\Delta(n)$ decreases in our simulations at each value of $s$.

\begin{figure}
\centering
\includegraphics[width=5.5cm]{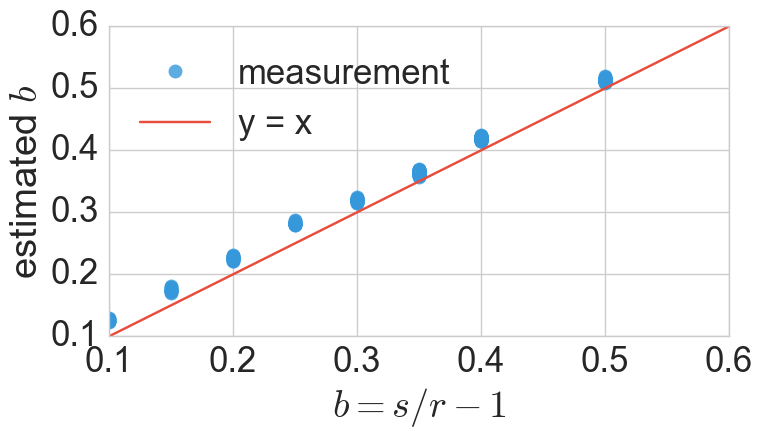}
\caption{Comparison of estimated and calculated $b$ values for Model I. Calculated values are based on the setting of $s$ and $r$, $b=s/r$-1. The estimated values are computed from the simulated average degree curves.}
\label{fig:model1-avgdeg-exponent}
\end{figure}

In Fig.~\ref{fig:model1-avgdeg-exponent}, we show that simulations fit well to the model equation~\eqref{eq:model1-avgdeg}.
We estimated $b$ from the observed function $\overline{d}(n)$ and compared that to $s/r-1$ that we calculated from the model parameters.
Our results indicate the growth of the average degree in our simulations follows~\eqref{eq:model1-avgdeg}.

\begin{figure}
\centering
\includegraphics[width=\columnwidth]{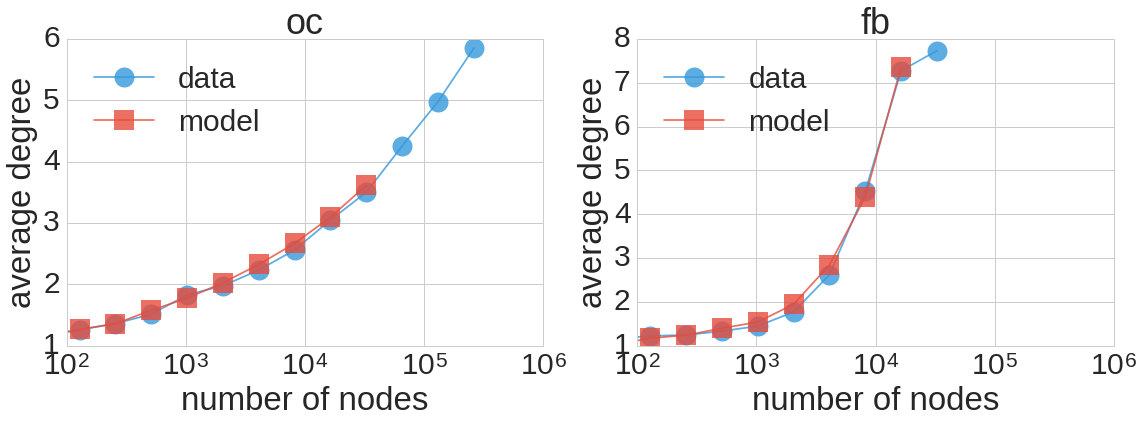}
\includegraphics[width=\columnwidth]{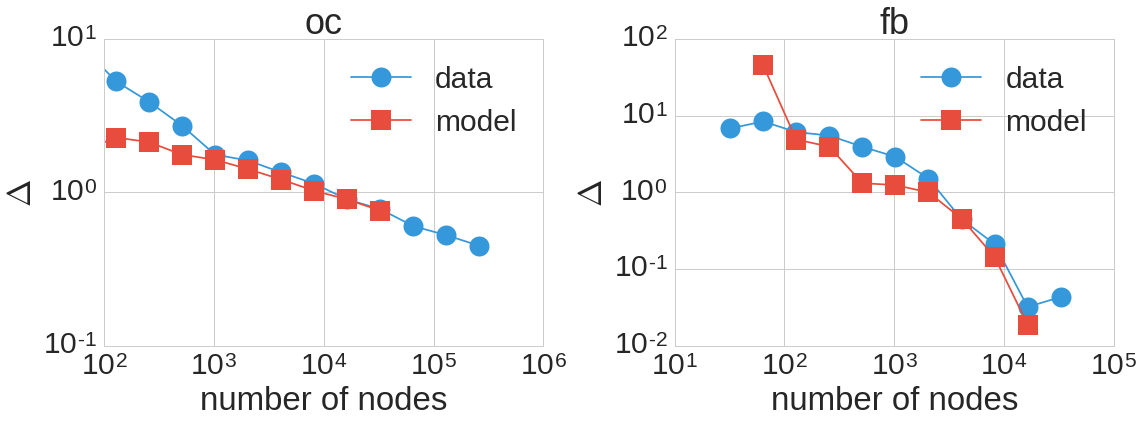}
\includegraphics[width=4.1cm]{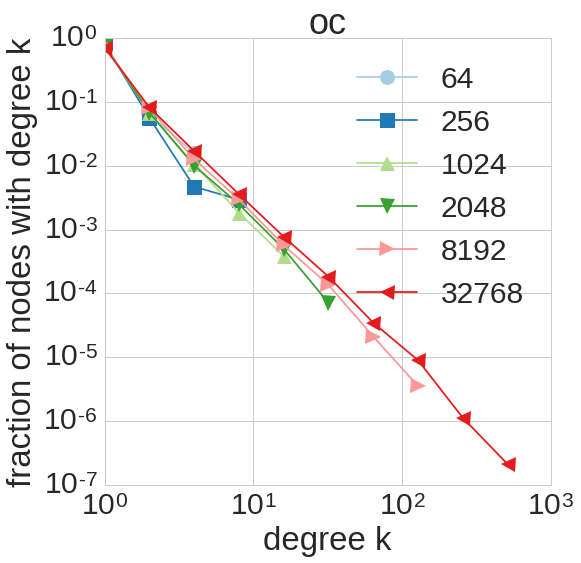}
\includegraphics[width=4.1cm]{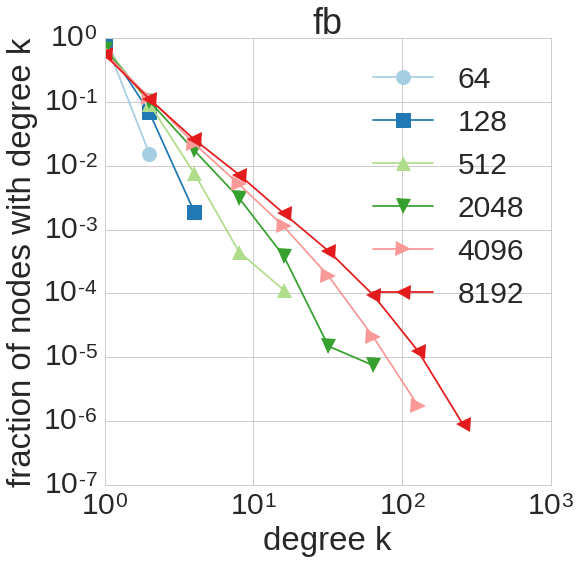}
\caption{Results of fitting the model to real data sets. \textbf{Left}: Occupy, \textbf{right:} Facebook. \textbf{Top}: Growth of the average degree. \textbf{Middle}: Decrease of the exponent of the degree distribution. \textbf{Bottom:} Visualization of the degree distributions at different network sizes.} 
\label{fig:model2-results}
\end{figure}

Next we turn to analyzing how Model~II fits to real networks.
We present the experiments for Occupy and Facebook; for the other networks, see Section~\ref{sect:reprod}.
For both data, we compute parameters $p,q,r,s,N_0$ and $H_0$ from the values of $a,b$ and $c$ in Table~\ref{tab:fit}.
Then we  analyze $\overline{d}(n)$ and $\Delta(n)$ in networks simulated by the corresponding parameters.
Specifically we set $p = 0.002$, $q = 0.022$, $r = 0.038$, $s =0.0645$, $N_0 = 14$, $H_0 = 2$ for Occupy, and $p = 0.0089$, $q = 0.0$, $r = 0.04$, $s=0.0857$, $N_0 = 85$, $H_0 = 2$ for Facebook.

In Fig.~\ref{fig:model2-results}, we show our measurements for the two data sets in separate columns.
By measuring in the real and the simulated data, we show the growth of $\overline{d}(n)$ on the top and of $\Delta(n)$ in the middle.
Finally, at the bottom, we include snapshots of the degree distributions of the simulated graphs.

We may conclude that by setting the proper values of the model parameters, our models are capable for reproducing the two main effects, increasing $\overline{d}(n)$ and decreasing $\Delta(n)$, as well as generating growing networks that fit well to our real-world data.

\section{Conclusion}

In growing networks, we measured that the exponent of the power-law degree distribution decreases in time.  We connected this observation with the growth of the average degree and gave models for this phenomenon.
In general, we model the initial growth of information networks by a random process as in an Erd\H{o}s-R\'enyi graph.
Then we observe that an organizing rule begins dominating the network building process, driving the network towards a scale-free degree distribution.
A model and its extended version that are based on exponential growth and preferential edges have been introduced in Section~\ref{sec:edge-models}.
This new model proposed in this paper yields a power-law degree distribution with decreasing exponent in a growing graph with increasing average degree.

In our future work we plan to extend our model to incorporate other network building mechanisms.
Most interestingly, we intend to incorporate triangle closing and vertex copying rules and analyze their effects on the degree distribution of the network.
We also plan to analytically derive the power law exponent as the function of the network size and the parameters of our models.

\section{Acknowledgments}

The publication was supported by ``Big Data---Momentum'' grant of the Hungarian Academy of Sciences and the PIAC\_13-1-2013-0197 project.

\bibliographystyle{abbrv}

\end{document}